\documentclass[12pt]{iopart}
\usepackage{graphicx}
\begin{document}

\title[Noise spectra in the reversible-irreversible transition...]{Noise spectra
in the reversible-irreversible transition in amorphous solids under oscillatory driving}

\author{Ido Regev$^{1}$, C Reichhardt$^{2}$ and C J O Reichhardt$^{3}$}

\address{$^{1}$Jacob Blaustein Institutes for Desert Research,
  Ben-Gurion University of the Negev,
Sede Boqer Campus 84990, Israel}
\address{$^2$Theoretical Division and Center for Nonlinear Studies,
Los Alamos National Laboratory, Los Alamos, New Mexico 87545,
United States of America
}
\ead{cjrx@lanl.gov}
\vspace{10pt}
\begin{indented}
\item[]\today
\end{indented}

\begin{abstract}
We study the stress fluctuations in simulations of a two-dimensional amorphous solid under a cyclic drive. It is known that this system organizes into a reversible state for small driving amplitudes and remains in an irreversible state for high driving amplitudes, and that a critical driving amplitude separates the two regimes. Here we study the time series of the stress fluctuations below and above the reversible-irreversible transition.  In the irreversible regime above the transition, the power spectrum of the stress fluctuations is broad and has a $1/f^{\alpha}$ shape with $1< \alpha <2$. We find that the low frequency noise power peaks near the stress at which dc yielding occurs, which is consistent with the behavior expected in systems undergoing a non-equilibrium phase transition.
\end{abstract}

%
\vspace{2pc}

\section{Introduction}

A transition from a non-fluctuating to a fluctuating
nonequilibrium state as a function of external driving appears
in many systems, including those that undergo
depinning transitions 
\cite{1}, collective transport \cite{2} jamming \cite{3}, and
static to active regime changes \cite{4,5}.
Critical features often appear at these transitions,
such as 
scaling behavior of the velocity-force curves, correlation functions, and structural
measures \cite{4,5}.
Similar effects appear near
the yielding transition under oscillatory shear,
where the system settles into a periodic state after some transient rearrangements
for driving amplitudes below yielding, but
can undergo continuous 
plastic rearrangements with large fluctuations for driving amplitudes above yielding \cite{6,7}.
It is often difficult to define the exact point at which yielding occurs using only the stress-strain curve; however,
a new method was recently proposed to identify the yielding transition
based on the reversible behavior of the system under a cyclic, rather than dc, drive \cite{8}.
This approach was inspired by work in dilute colloidal systems under periodic shear \cite{9}, in which colloid-colloid collisions die away at low shear and
the system organizes to a completely reversible state
after a number of shear cycles that diverges at a critical maximal shearing amplitude \cite{10}.
Above the critical shearing amplitude, 
the system remains in an irreversible state.
In a similar study on a jammed amorphous system,
at small ac shear amplitudes the sample initially behaves irreversibly from
one cycle to the next, but after a transient period it settles into a reversible state in
which all particles return to the same positions at the end of each shear cycle.
As the shear amplitude increases, the number of cycles needed to converge to
a reversible state diverges near a critical drive amplitude \cite{8,11,31}.
Above the transition, the system
remains in an irreversible state,
while the critical maximal amplitude of the cyclic driving at the onset of 
irreversibility is close to the expected yielding point of the stress-strain curve under a dc drive \cite{8,12}.
  
Time series analysis methods, such as the power spectrum of noise fluctuations, can be used to
identify and characterize changes in the dynamics of driven many-body systems.
Such methods have been applied
to sliding charge density waves \cite{25},
driven superconducting vortices \cite{26,27},
and Barkhausen noise in magnetic systems \cite{28}.
The use of these techniques for fluctuations 
in systems that exhibit yielding transitions
is problematic since there are typically no fluctuations below the yielding
transition.
Subjecting the system to a cyclic drive induces fluctuations even in
the reversible state below
yielding, making a power spectrum analysis possible.
Here we show that the spectra contain
useful information
about the dynamics of
the ac driven yielding system both above and below the yield transition.

\section{Simulation and system}

\begin{figure}
\includegraphics[width=0.8\columnwidth]{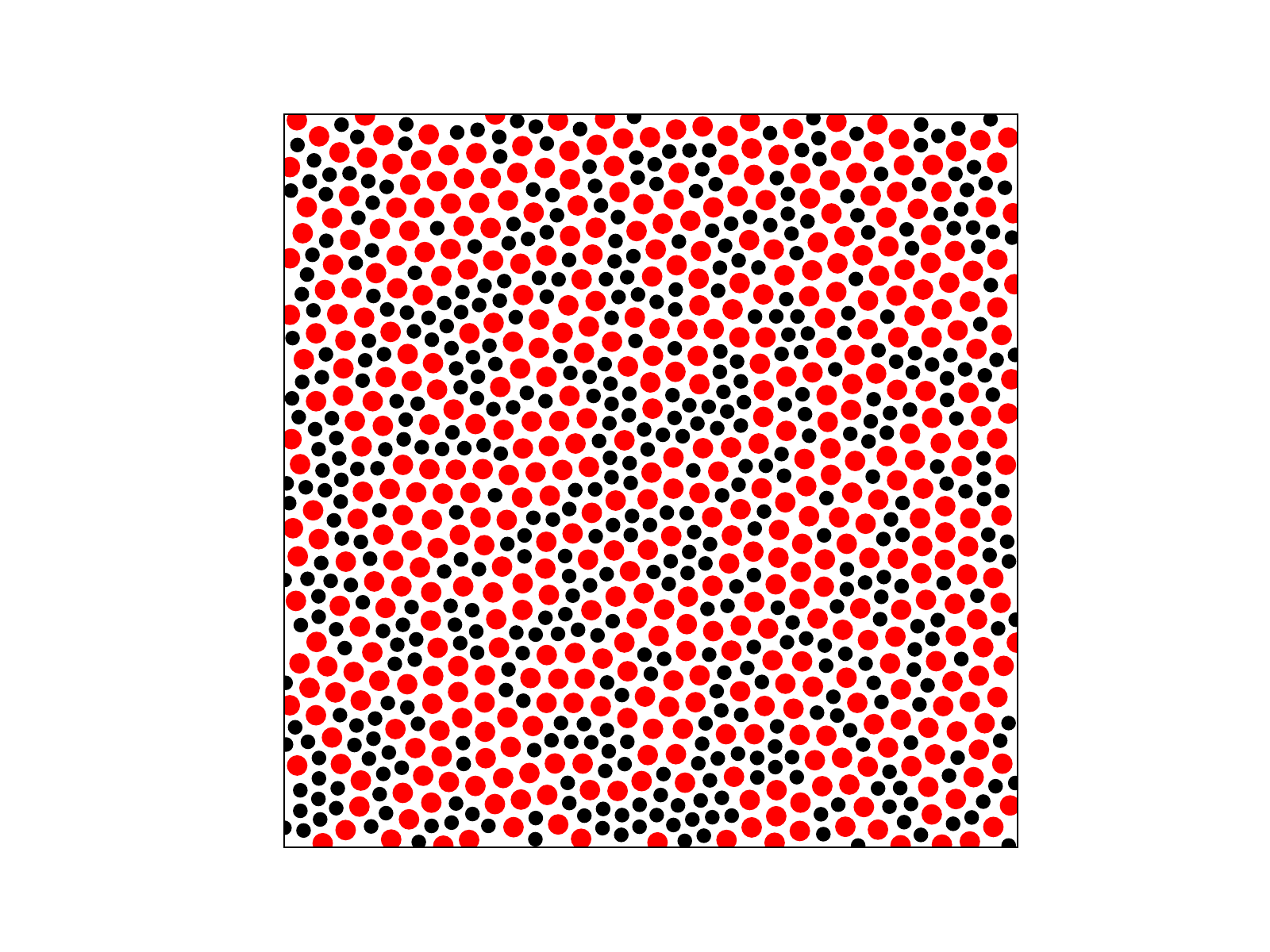}
\caption{
  A typical snapshot of the simulated system containing
  a bidisperse assembly of particles with a radius size ratio of 1.0 to 1.4 in an
  amorphous configuration above the jamming density.
  When subject to a periodic strain, the particles either organize into
  a reversible state or keep diffusing irreversibly, depending on the maximal strain amplitude.
}
\label{fig:1}
\end{figure}

We consider a two-dimensional amorphous system  
containing a bidisperse assembly of pairwise interacting particles.
One particle species
has a radius 1.4 times larger than
that of the other species,
so the system forms a disordered state.
We fix the density of the particles at $\phi=0.75$, which is higher than the
jamming density.  The initial state is obtained by quenching the system from a
high temperature down to $T=0$ in order to obtain an amorphous configuration.
The particles move according to overdamped dynamics as described in
detail in \cite{old14} and \cite{8}.
We apply a periodic athermal quasi-static shear with maximal
strain amplitude $A$ in each realization.
In figure~\ref{fig:1} we show a snapshot of an initialized system
with a disordered structure.
For more details of the simulations, see \cite{8}.

\section{Results and discussion}

\begin{figure}
\includegraphics[width=0.8\columnwidth]{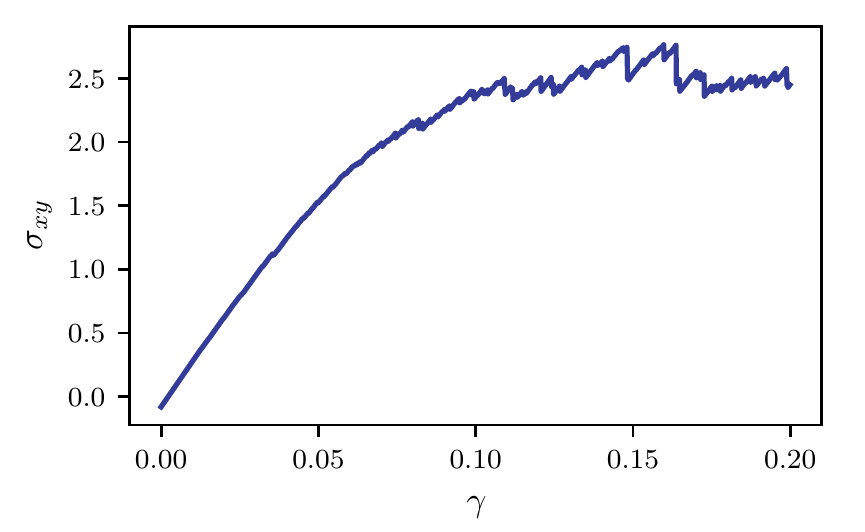}
  \caption{The dc stress-strain curve $\sigma_{xy}$ vs $\gamma$ for the system in
    figure~\ref{fig:1}
showing the onset of of large fluctuations near $\gamma > 1.0$.  
}
\label{fig:2}
\end{figure}

In figure~\ref{fig:2} we plot a typical dc stress ($\sigma_{xy}$) versus
strain ($\gamma$) curve.
At low strains,
$\sigma_{xy}$ increases with
increasing strain until it begins to saturate and starts to 
exhibit significant fluctuations at $\gamma \approx 0.9$. If we keep increasing the strain, the stress continues to fluctuate forever in a chaotic non-repetitive manner.
If, however, we apply a cyclic or ac drive with a maximal strain amplitude $\gamma$,
we observe two different types of behavior.  When $\gamma$
is smaller than the dc yielding transition, the system organizes into a reversible state
in which the stress fluctuations are still noisy but repeat exactly after each shearing cycle.
When $\gamma$ is larger than the dc yielding transition,
we obtain chaotic, non-repetitive dynamics.

\begin{figure}
\includegraphics[width=0.8\columnwidth]{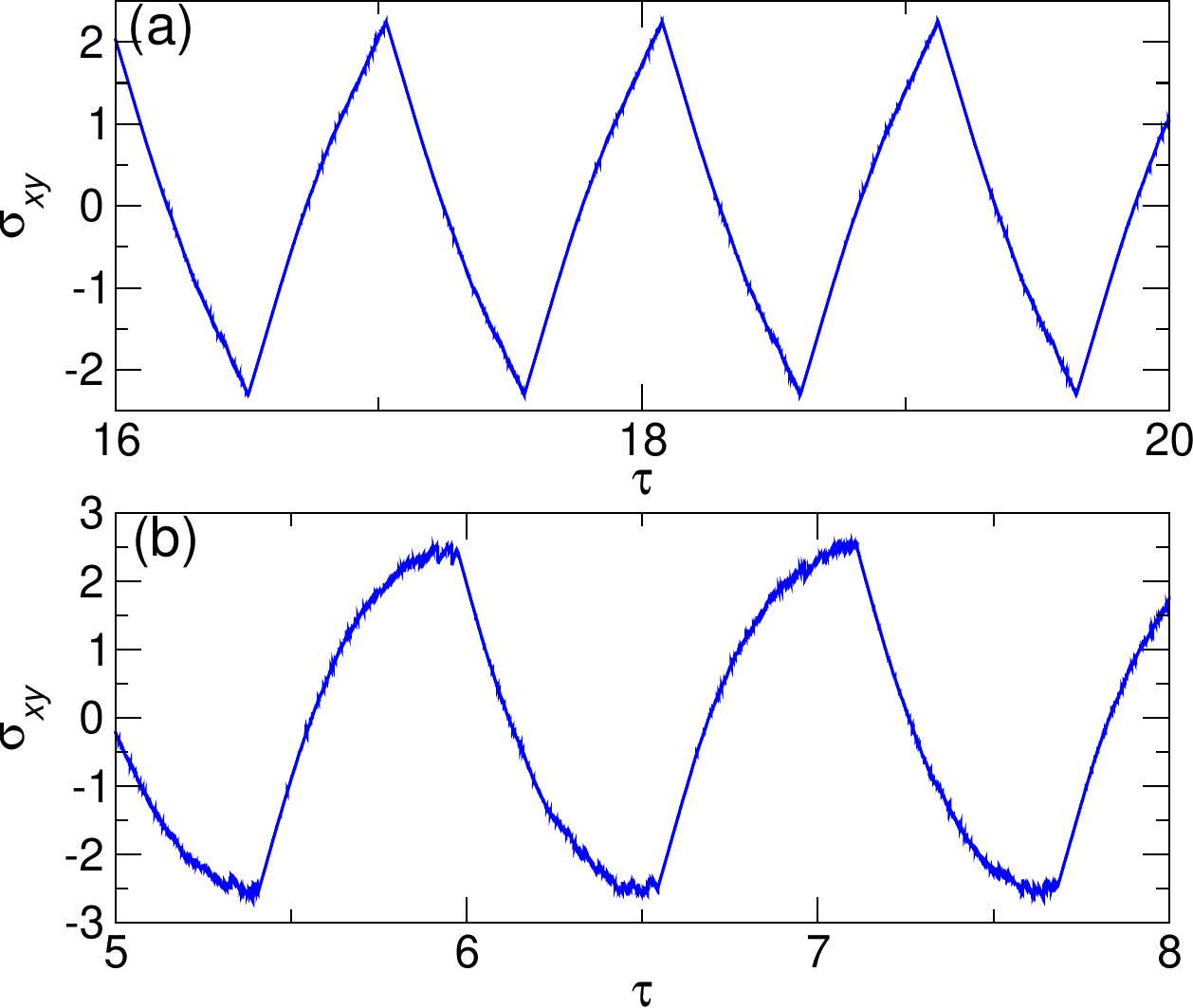}
\caption{ A portion of the time series of $\sigma_{xy}$ versus time $\tau$ in ac cycle
  periods
  for the system in figure~\ref{fig:2} under a periodic drive.
  (a) At $\gamma = 0.08$, the
system is in a reversible state.
(b) At $\gamma = 0.15$, the system is in a irreversible state.  The largest fluctuations
in $\sigma_{xy}$ occur near the extremal values of $\sigma_{xy}$, but the shape
of the fluctuations differs from cycle to cycle.
}
\label{fig:3}
\end{figure}

\begin{figure}
  \includegraphics[width=0.8\columnwidth]{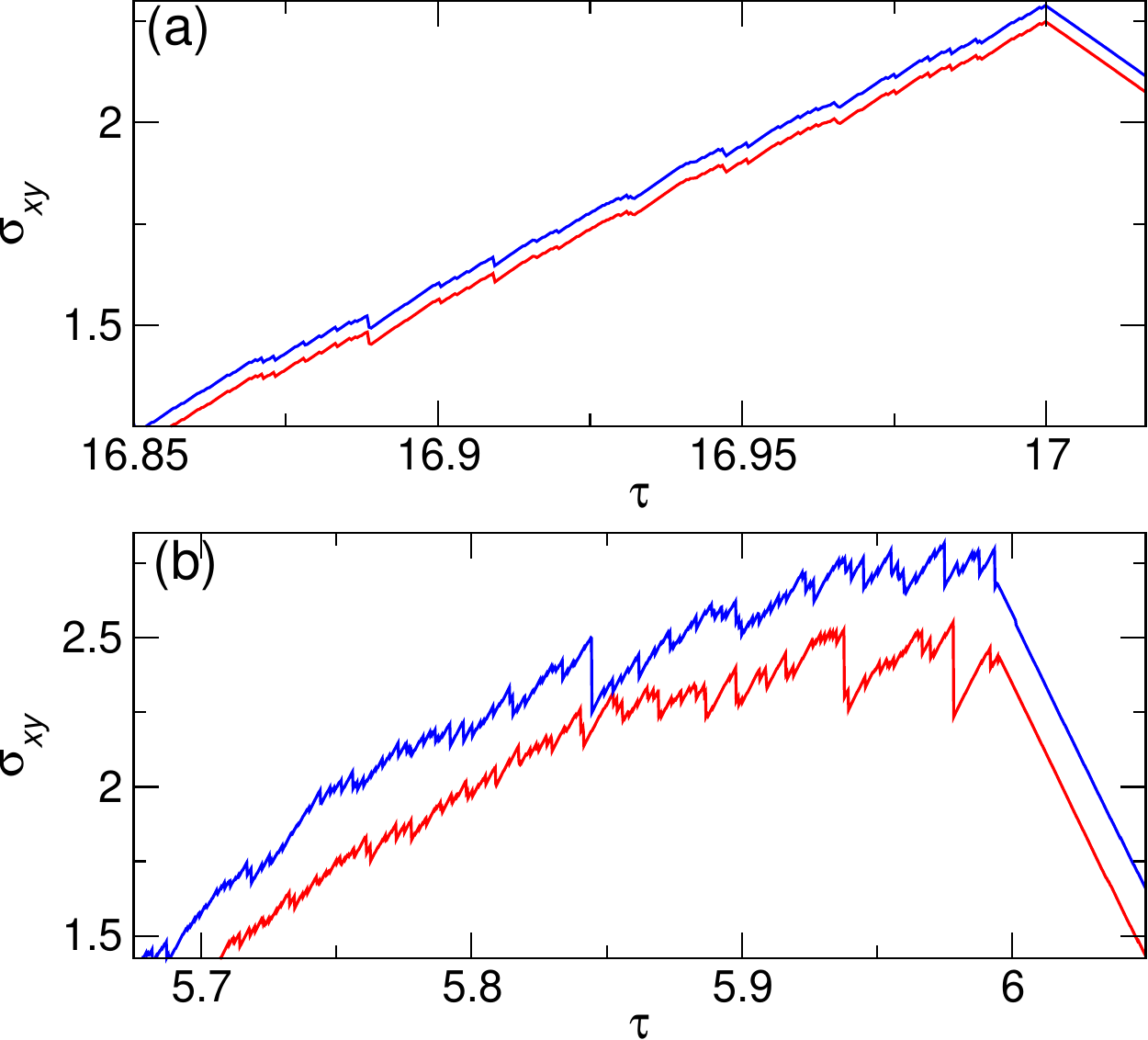}
  \caption{Zoomed-in portions of the time series of $\sigma_{xy}$ versus $\tau$
    in ac cycle periods from figure~\ref{fig:3}.  Red curves are from cycle $n$ and
    blue curves, which are offset slightly in the $y$ direction, are from cycle
    $n+1$.  (a) At $\gamma=0.08$ in the reversible state, the detailed noise on
    each cycle is repeated.  (b) At $\gamma=0.15$ in the irreversible state, the
    detailed noise differs from cycle to cycle.
  }
  \label{fig:3A}
\end{figure}

In figure~\ref{fig:3}(a)
we show a time series of $\sigma_{xy}$ in the reversible regime 
at $\gamma = 0.08$, where
we find 
a smooth oscillatory curve with the same noise pattern appearing in each drive cycle,
as illustrated in figure~\ref{fig:3A}(a).
In figure~\ref{fig:3}(b) at
$\gamma = 0.15$, in the irreversible regime,
there are
larger fluctuations
near the maxima and minima in $\sigma_{xy}$, and the noise fluctuations are
different from one cycle to the next, as shown in figure~\ref{fig:3A}(b).

\begin{figure}
\includegraphics[width=0.8\columnwidth]{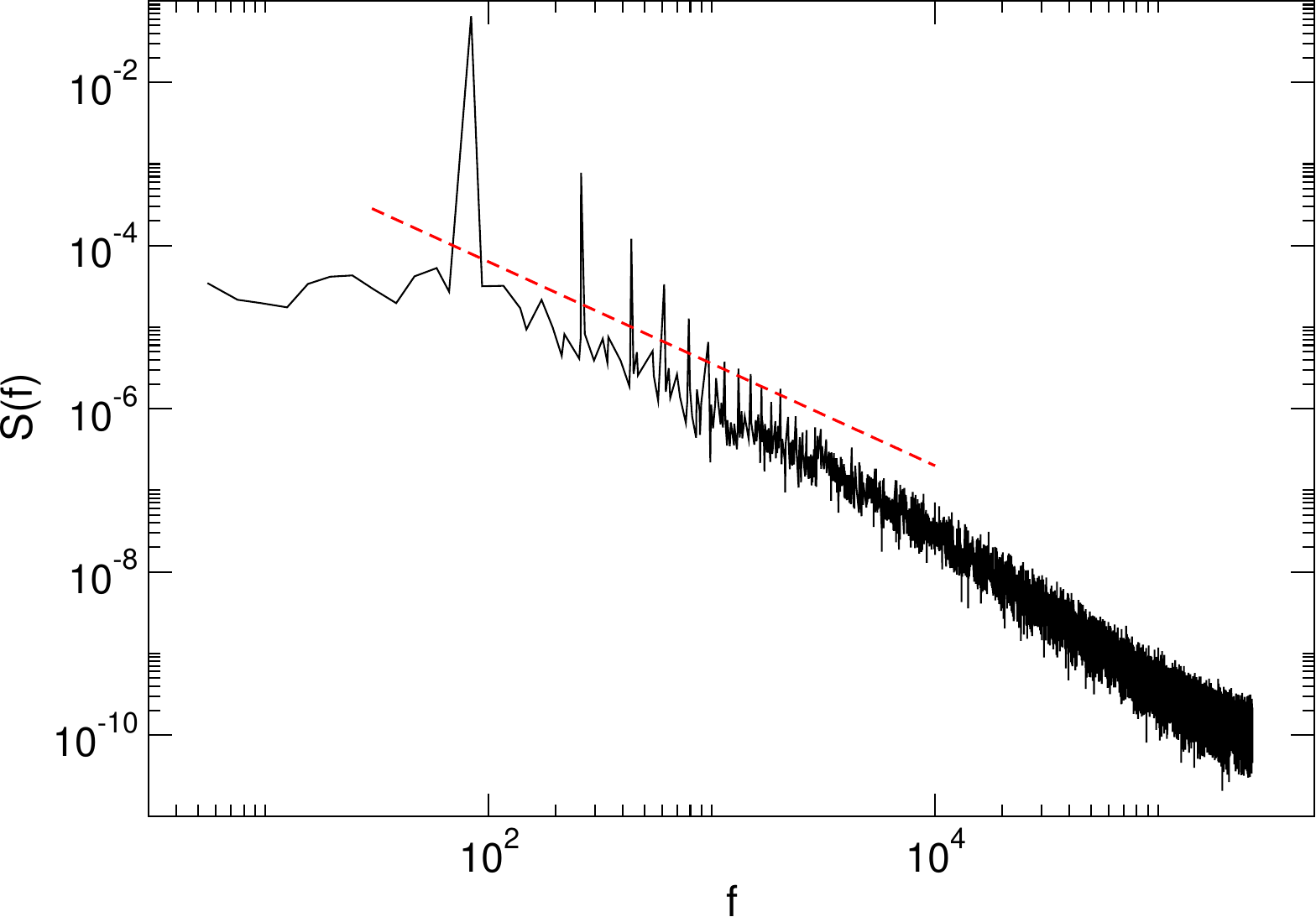}
\caption{The power spectrum $S(f)$ for the system in
  figure~\ref{fig:3}(b) at $\gamma = 0.15$ in the irreversible 
  regime.
  The ac driving frequency appears as a series of peaks superimposed on
  a $1/f^{\alpha}$ background with
  $\alpha \approx 1.25$.  The dashed line is a power law fit with exponent 1.25.
  At higher frequencies, the system has
a $1/f^2$ behavior. 
}
\label{fig:4}
\end{figure}

We next examine the power spectrum of the time series, 
\begin{equation}
S(\omega) =\frac{1}{2\pi}\left |\int \sigma_{xy}(t)e^{i\omega t} dt \right|^2 ,
\end{equation}
where the frequency $f = \omega/2\pi$. 
In figure~\ref{fig:4} we plot the power spectrum $S(f)$ versus $f$ 
of the time series taken from 100
ac drive cycles for the system in
figure~\ref{fig:3}(b) at a strain amplitude of $\gamma = 0.15$. 
The large spike in 
$S(f)$ is at the ac driving frequency, and the smaller spikes are the harmonics
of this frequency.
Between the spikes the curve is relativity  smooth, indicating a
lack of response at other specific frequencies due to the irreversible, non-periodic, nature of the dynamics.
The dashed line indicates a fit to
$S(f) \propto f^{-\alpha}$ with $\alpha = 1.25$.
This
resembles
the $1/f^{\alpha}$
noise
frequently observed
in strongly fluctuating driven disordered systems such as
superconducting vortices
or domain walls exhibiting Barkhausen noise, where
$1.0 < \alpha < 2.0$ \cite{1}. We note that the behavior at
high frequencies is better fit by $\alpha=2$;
however, the large scale avalanches and
plastic rearrangements that occur in the system
are associated with much slower time scales corresponding to the
low frequency dynamics.

\begin{figure}
\includegraphics[width=0.8\columnwidth]{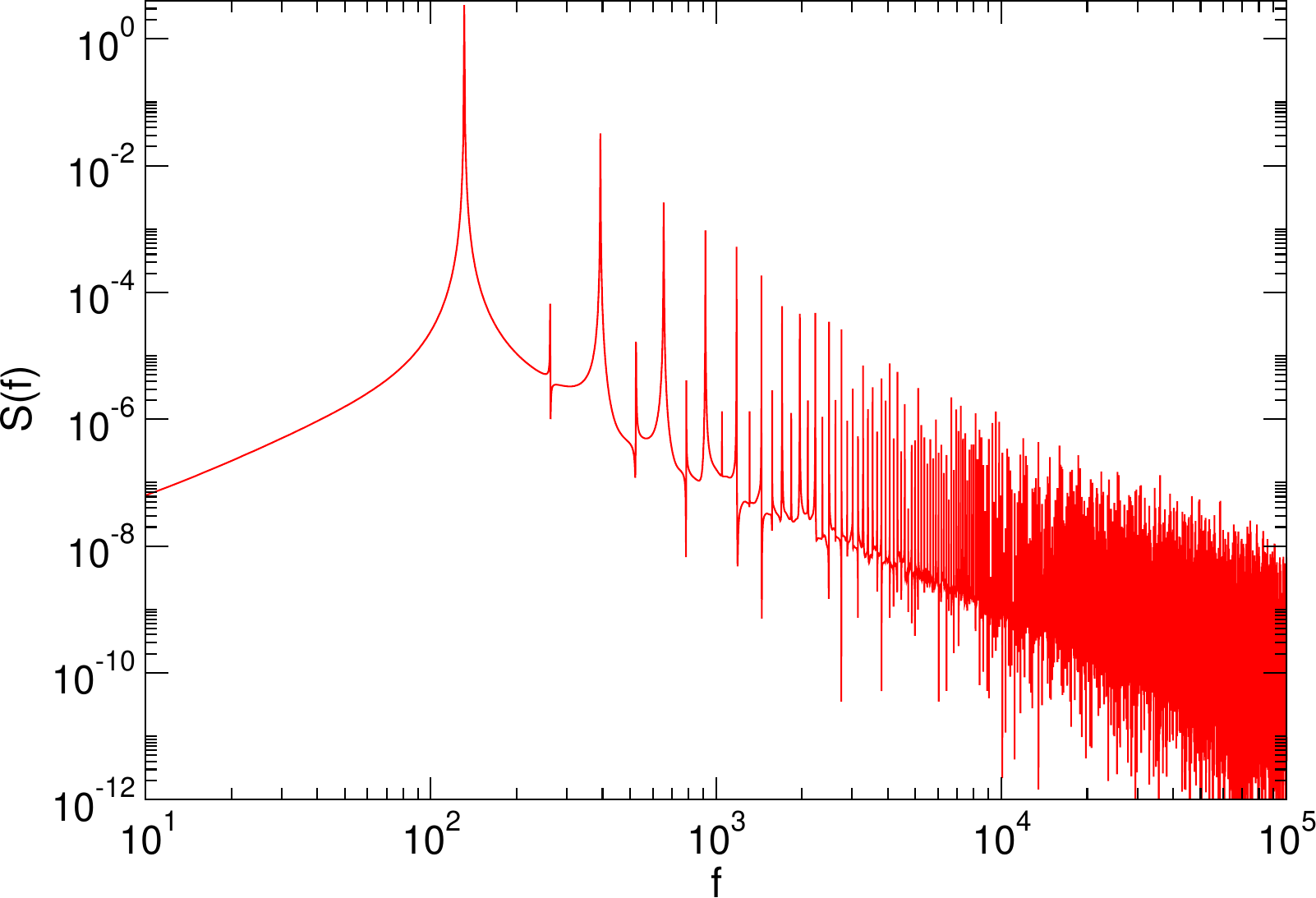}
\caption{(a) The power spectrum $S(f)$
  in the reversible regime at $\gamma = 0.08$, where a large number of additional
peaks appear in addition to those associated with the driving frequency. 
}
\label{fig:5}
\end{figure}

In figure~\ref{fig:5}(a) we plot $S(f)$ for the same system at  $\gamma = 0.08$ 
where the behavior is reversible.
Here we find numerous spikes in the power spectrum
in addition to those associated with the ac driving frequency.
These additional spikes are produced by the
repeatable avalanches or plastic events
that each have their own 
characteristic frequency,
so that the overall power spectrum 
is best described as a collection of specific frequencies rather than
as $1/f^{\alpha}$ noise. 
In general, as $\gamma$ increases in the reversible regime, the number of
additional power spectrum spikes increases since a larger number of distinct
reversible events occur,
and when the system reaches the irreversible regime,
a true broad band background noise signature emerges.

\begin{figure}
\includegraphics[width=0.8\columnwidth]{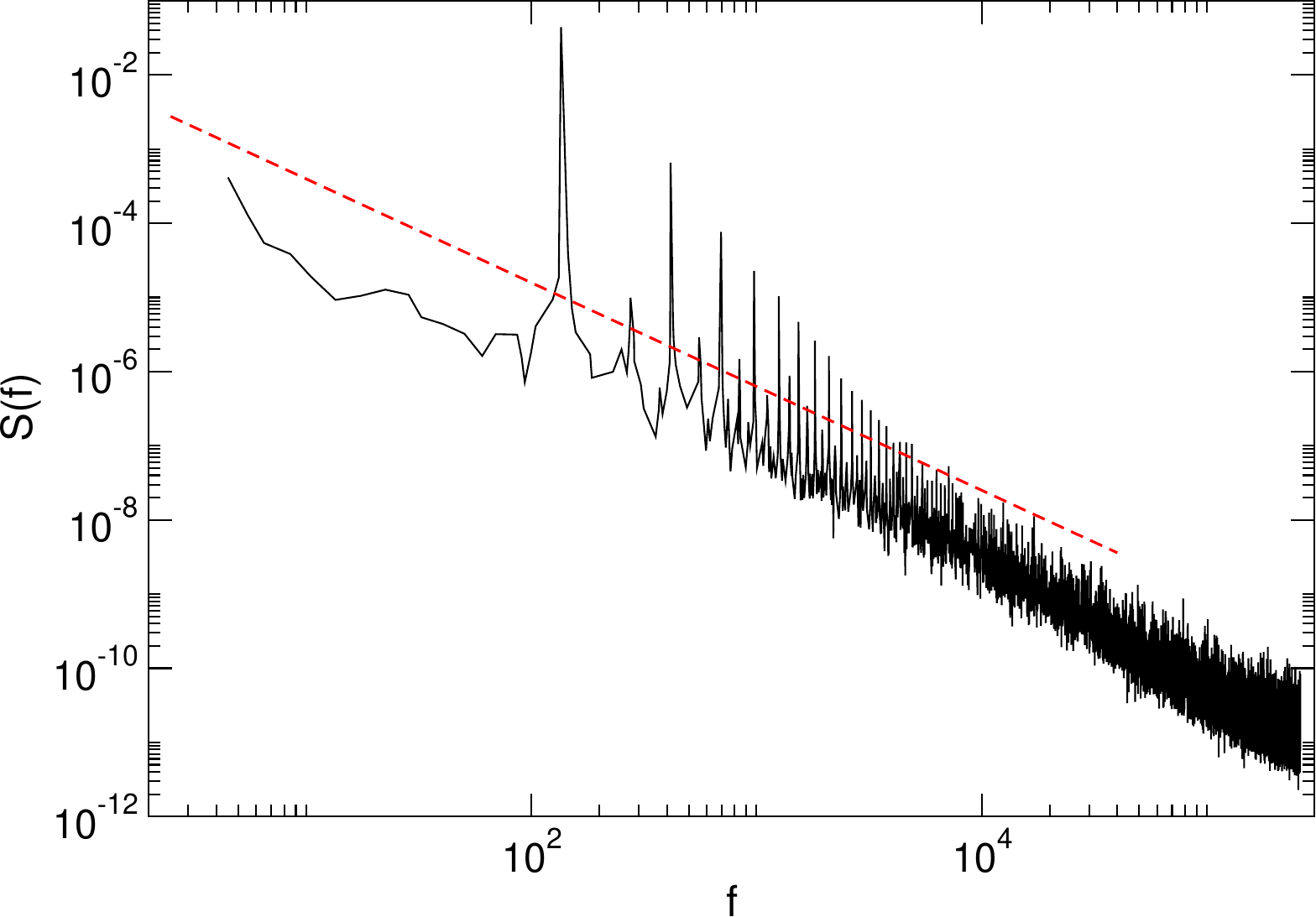}
\caption{ The power spectrum $S(f)$
  for $\gamma = 0.093$, close to the yield point. 
  Here the lower frequency power is much higher than what is observed
  at $\gamma = 0.15$ in figure~\ref{fig:4}, indicating the appearance of
  slower, large scale rearrangements.
  The dashed line indicates a power law fit with $\alpha = 1.35$,
  and the fit extends over a wider range of frequencies
  compared to the $\gamma = 0.15$ behavior.  
}
\label{fig:6}
\end{figure}

In figure~\ref{fig:6} we plot $S(f)$
at $\gamma = 0.93$, closer to the yielding threshold.
Compared to the $\gamma=0.15$ spectrum, here we find
that the peaks produced by the ac drive are smaller since the
particle arrangements become more random as
the yielding transition is approached.
At the same time, the low frequency power for $f<100$ is higher
at $\gamma=0.93$, indicating the occurrence of large scale, and thus
long time, rearrangements in the system.
Previous work on the noise power of fluctuating quantities near a
critical point indicates that the largest low frequency noise occurs close
to the transition \cite{29,30}.  The fluctuations are small and local both well below the
critical point and well above it, but near the critical point the fluctuations become
self-similar and extend across all sizes, giving a maximum in the low frequency
noise power.
The dashed line in figure~\ref{fig:6} indicates a power law fit to
$1/f^{\alpha}$ with $\alpha =1.35$.   The power law behavior
extends over a larger range of frequencies than for the
$\gamma = 0.15$ system in figure~\ref{fig:4},
and the power law exponent $\alpha$ is somewhat larger than in the $\gamma=0.15$
case.

\begin{figure}
\includegraphics[width=0.8\columnwidth]{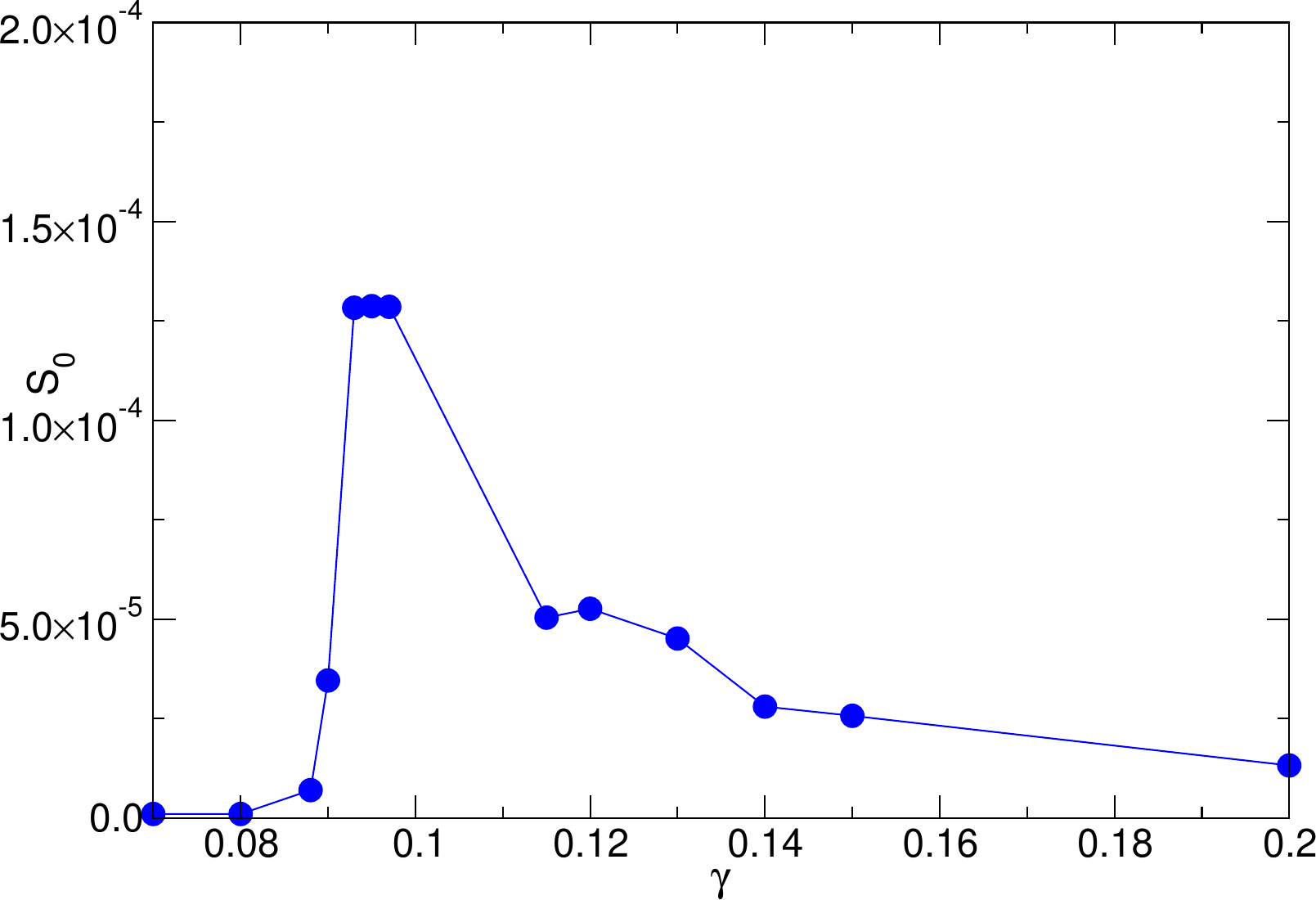}
\caption{ 
The noise power $S_{0}$ at the lowest measured frequencies from the power spectra vs $\gamma$ 
showing that the low power noise is largest near $\gamma = 0.1$ close to the dc yielding
point observed in figure~\ref{fig:2}. 
}
\label{fig:7}
\end{figure}

In figure~\ref{fig:7} we plot the noise power $S_{0}$ at the
lowest frequencies versus $\gamma$.
The noise power is largest 
near $\gamma = 0.1$, close to the yielding point,
and it drops off for higher $\gamma$.
We note that
for $\gamma < 0.07$, $S_{0}$ is close to zero since
all the fluctuations are lost and the system
begins to behave elastically. 
A peak in the noise power
is expected near a critical point,
and in this same regime,  the exponent of
the power spectrum can contain information about the
universality class \cite{30}.
A more careful analysis would be needed to find the exact critical point
and exponents; however,
our results clearly show that analyzing the power spectra can be a useful tool for understanding the
reversible-irreversible transition in periodically driven amorphous systems. 

\section{Summary}
We have examined the transition from reversible to irreversible dynamics in a
periodically shared
two dimensional amorphous system by
measuring the power spectra of the time series of
the stress versus strain.
For maximal driving amplitudes $\gamma$, for which the system is reversible, the power spectrum shows a peak at the ac drive frequency along with
a series of smaller peaks associated with reversible plastic events that repeat with typical frequencies.
In the irreversible regime, the only peaks present are associated with the driving frequency, and the peaks appear
on top of a $1/f^{\alpha}$ background noise response, where $1.0 < \alpha < 1.5$.
We find that the low frequency noise power is maximal near the dc yielding point, a behavior similar to that observed
near other nonequilibrium transitions.
Our results show that features of the power spectra can be useful in characterizing reversible-irreversible transitions in amorphous systems.  
Further directions for study include filtering out the driving frequency in order to focus exclusively on the periodic rearrangements as well as
adding thermal noise or other fluctuations to examine the robustness of the peak in the noise power near the transition.

\ack
We gratefully acknowledge the support of the U.S. Department of
Energy through the LANL/LDRD program for this work.
This work was supported by the US DoE through Los Alamos
National Laboratory.  Los Alamos National Laboratory is operated
by Triad National Security, LLC, for 
the 
NNSA of the 
U.S. DoE
(Contract No. 892333218NCA000001).
I.R. would like to thank the Israel Science Foundation (ISF) for its support through Grant No. 1301/17.

\end{document}